\documentclass[preprint,12pt]{elsarticle}

\usepackage[margin=1in]{geometry}

\usepackage{graphicx}

\usepackage{amssymb}

\usepackage{url}

\usepackage{amsmath}
\usepackage{upgreek}

\usepackage{amsthm}

\def\doubleunderline#1{\underline{\underline{#1}}}
\usepackage{tensor}
\usepackage{dsfont}

\usepackage{siunitx}

\usepackage{amssymb}

\usepackage{bbm}

\usepackage{soul}

\usepackage{chemformula}

\usepackage{algorithmicx}
\usepackage{algorithm}
\usepackage[noend]{algpseudocode}
\makeatletter
\renewcommand{\ALG@beginalgorithmic}{\footnotesize}
\makeatother

\usepackage{booktabs}

\usepackage{amsmath}

\makeatletter
\newcommand{\vast}{\bBigg@{4}}
\newcommand{\Vast}{\bBigg@{7}}
\makeatother

\makeatletter
\def\ps@pprintTitle{%
  \let\@oddhead\@empty
  \let\@evenhead\@empty
  \let\@oddfoot\@empty
  \let\@evenfoot\@oddfoot
}
\makeatother

\usepackage{todonotes}

\newcommand*\xbar[1]{%
   \hbox{%
     \vbox{%
       \hrule height 0.5pt 
       \kern0.5ex
       \hbox{%
         \kern-0.1em
         \ensuremath{#1}%
         \kern-0.1em
       }%
     }%
   }%
}

\usepackage{caption}

\usepackage{subcaption}

\setcitestyle{authoryear}
 \biboptions{comma,round}

\date{}

\journal{Elsevier}

\begin{document}

\begin{frontmatter}

\title{A model for transport of interface-confined scalars and insoluble surfactants in two-phase flows}


\author{Suhas S. Jain\corref{cor1}}
\ead{sjsuresh@stanford.edu}

\address{Center for Turbulence Research, Stanford University, California, USA 94305}

\begin{abstract}

In this work, we propose a novel scalar-transport model for the simulation of scalar quantities that are confined to the interface in two-phase flows. In a two-phase flow, the scalar quantities, such as salts and surfactants, can reside at the interface (due to their molecular structure, electrostatic interactions, and solubility) and can modify the properties of the interface, in the time scales of interest. This confinement of the scalars leads to the formation of sharp gradients of the scalar concentration values at the interface, presenting a serious challenge for its numerical simulations.

To overcome this challenge, we propose a computational model for the transport of scalars that maintains the confinement condition for these quantities. The model is discretized using a central-difference scheme, which leads to a non-dissipative implementation that is crucial for the simulation of turbulent flows. The model is used with the ACDI diffuse-interface method \citep{jain2022accurate}, but can also be used with other algebraic-based interface-capturing methods. Furthermore, the provable strengths of the proposed model are: (a) the model maintains the positivity property of the scalar concentration field, a physical realizability requirement for the simulation of scalars, when the proposed criterion is satisfied, (b) the proposed model is such that the transport of the scalar concentration field is consistent with the transport of the volume fraction field, which results in effective discrete confinement of the scalar at the interface; and therefore, prevents the artificial numerical diffusion of the scalar into the bulk region of the two phases.

Finally, we present numerical simulations using the proposed model for both one-dimensional and multidimensional cases and assess: the accuracy and robustness of the model, the validity of the positivity property of the scalar concentration field, and the confinement of the scalar at the interface.

\end{abstract}


\end{frontmatter}



\section{Introduction} 


The transport of scalars on evolving interfaces in fluids is an ubiquitous phenomenon across a broad range of processes in nature and in engineering, and is worthy of modeling. 
We, here, refer to an interface-confined scalar as any passive or active scalar quantity that is transported along an evolving/deforming interface. 
These scalars could represent charged species \citep{chu2007surface}, surfactants \citep{hargreaves2007chemical}, or any other conserved scalar quantity.



Surfactants lower the surface tension properties and generate Marangoni forces, which are useful in controlling the dynamics of multiphase flows. They are transported with the interface due to convection, and they diffuse along the interface when there is a concentration gradient, and can also be exchanged (adsorbed/desorbed) between the bulk and the interface \citep{defay1966surface}. Surfactants have applications in industry in emulsification and mixing, drug delivery, droplet manipulation in microfluidics \citep{eggleton2001tip,booty2005steady,baret2012surfactants,pit2015droplet}, drag reduction \citep{manfield1999drag}, and are important for the functioning of lungs \citep{yap1998influence}.









Research on modeling surfactants dates back to \cite{stone1990simple} and \cite{wong1996surfactant}. 
The effects of surfactants on drop/bubble deformation and breakup have been studied theoretically or semi-analytically by \citet{stone1990effects,milliken1993effect,milliken1994influence,pawar1996marangoni,siegel1999influence}. One of the first coupled numerical simulations studying the effect of surfactants on the flow around bubbles was done by \citet{cuenot1997effects}. Since then, various methods have been proposed for modeling insoluble surfactants. 
Using a boundary integral method, \citet{li1997effect,yon1998finite,eggleton1999insoluble} studied the effect of insoluble surfactants on drops in a Stokes flow, and \citet{eggleton2001tip} simulated the tip streaming breakup of drops that occurs in the presence of surfactants. A coupled grid-based particle method with an implicit boundary integral method was also proposed by \cite{hsu2019coupled}.


In continuum approaches, and in the context of sharp-interface methods, insoluble surfactants have been modeled using a volume-of-fluid method by \cite{renardy2002new,drumright2004effect,james2004surfactant}; 
using a Lagrangian-based finite-element formulation by \cite{pozrikidis2004finite,ganesan2009coupled,venkatesan2019simulation,frachon2023cut}; 
using a segment projection method by \cite{khatri2011numerical};
using a level-set method by \cite{xu2003eulerian,xu2006level,xu2012level};
using a front-tracking method with adaptive mesh refinement by \cite{de20153d};
using an immersed-boundary method by \cite{lai2008immersed};
and using hybrid methods by \cite{ceniceros2003effects,cui2011computational}.

In the context of diffuse-interface methods, a free energy functional-based model has been used to model surfactants by \cite{van2006diffuse,yun2014new}, and the well-posedness of the system has been studied by \cite{engblom2013diffuse,abels2019existence,di2022well}.
All the existing methods in the literature use a Cahn-Hilliard-based diffuse-interface framework for modeling surfactants \citep{teigen2009diffuse,teigen2011diffuse,garcke2014diffuse,ray2021discontinuous}.
More recently, the effects of surfactants on breakup and coalescence of droplets in a turbulent flow, was studied by \cite{soligo2019breakage}, along with their feedback effect on the flow by \cite{soligo2020effect}. However, to the best of our knowledge, there is no model for transport of surfactants or interface-confined scalars for second-order phase-field methods. 

We recently developed a model for transport of scalars in the bulk of one of the phases in a two-phase flow \citep{jain2023scalar} given by
\begin{equation}
\frac{\partial c}{\partial t} + \vec{\nabla}\cdot(\vec{u} c) = \vec{\nabla} \cdot \left[D \left\{\vec{\nabla}c - \frac{(1 - \phi) \vec{n} c}{\epsilon} \right\}\right],
\label{equ:bulk_scalar}
\end{equation}
where $c$ is the scalar concentration (amount of scalar per unit volume), and showed that the model results in the consistent transport of scalar with the phase-field variable and will not result in artificial leakage of scalar across the interface.  
This model was also extended to include transfer across the interface by \citet*{mirjalili2022computational}.

The primary objective of the present work is to propose a computational model for transport of scalars, where the scalars are confined to the interfacial region while they are allowed to evolve along the interface. In this work, we propose a consistent method that results in leakage-proof transport of scalars along the convecting and deforming material interface. 
The proposed model does not require division by $\phi$ (volume fraction), and therefore doesn't require any special treatment when $\phi$ goes to $0$.  
We also prove and show that the scalar concentration value remains positive, which is a physical-realizability condition, using second-order central-difference schemes.

We use a second-order phase-field method, particularly the accurate conservative phase-field/diffuse-interface (ACDI) method by \cite{jain2022accurate}, for modeling the interface in a two-phase flow. 
The proposed interface-confined scalar model in this work can also be used with a conservative phase-field/diffuse-interface (CDI) method \citep{chiu2011conservative}, 
a conservative level-set (CLS) method \citep{olsson2005conservative}, an accurate conservative level-set (ACLS) method \citep{Desjardins2008}, including the five-equation and four-equation compressible diffuse-interface methods in \citet{jain2020conservative,jain2023assessment}, and any other method that results in a hyperbolic tangent interface shape in equilibrium, and when the volume fraction $\phi$ is bounded between $0$ and $1$. For coupling with other models, like a Cahn-Hilliard model where the volume fraction takes values between $-1$ and $1$, the proposed model can be affine transformed with respect to the phase-field parameter, such that the change in the range from $[0,1]$ to the range of values of $\phi$ that the interface-capturing model admits is accounted for.  
 
We present simulations of transport of scalars confined to the interface to illustrate the accuracy, consistency, and robustness of the proposed method in localizing the scalar to the interface location and in maintaining the leak-proof condition.  

\subsection{Governing equations: sharp and diffuse representations}

Consider the schematic of an interface $\gamma$ in a domain $\Omega$ shown in Figure \ref{fig:sharp_flux} along with the sharp and diffused representation of an interface-confined scalar or surfactant. Physically, when a scalar/surfactant is confined to the interface region, the concentration can be mostly represented as a sharp quantity for macroscopic continuum modeling. This is because the thickness of the scalar/surfactant layer, that is adsorbed onto the interface, is typically on the $O(nm)$ thick. 

However, this sharp nature of the scalar concentration poses a challenge in the numerical modeling of these scalars on an Eulerian grid because of the sharp jump in the concentration values. To overcome this issue, once could artificially diffuse the scalar in the interface normal direction as shown in Figure \ref{fig:sharp_flux} (right) in such a way that the gradients in the concentration can now be resolved on an Eulerian grid. If $\hat{c}$ represents the scalar/surfactant concentration per unit area in the sharp representation, then we could construct a diffuse quantity $c$ (concentration per unit volume) in such a way that the integral of this quantity in the interface normal coordinate will result in $\hat{c}$. Hence, if $\hat{c}$ represents the concentration of a conserved quantity, $c$ will also be a conserved variable. 

\begin{figure}
    \centering
    \includegraphics[width=0.75\textwidth]{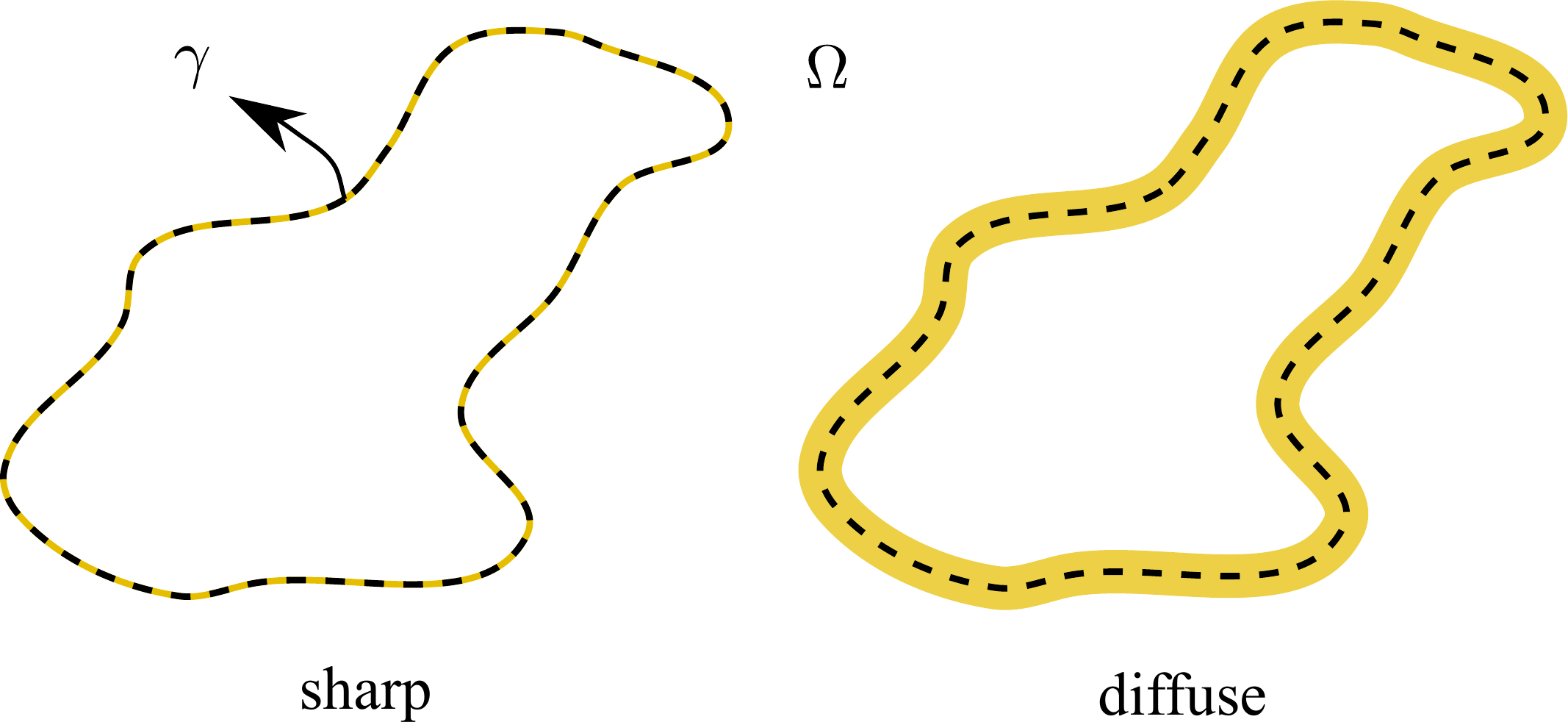}
    \caption{Schematic representing interface-confined scalars in sharp and diffuse representations. Here, $\gamma$ represents the two-dimensional interface embedded in a three-dimensional domain $\Omega$. The dashed line represents the interface, and the colored solid line represents an interface-confined scalar/surfactant.}
    \label{fig:sharp_flux}
\end{figure}


The evolution equation for surfactants on the interface (in a sharp representation) can be written as \citep{stone1990simple,wong1996surfactant}
\begin{equation}
\frac{\partial \hat{c}}{\partial t} + \vec{u}_s\cdot \vec{\nabla}_s \hat{c} = \vec{\nabla}_s \cdot \left(D \vec{\nabla}_s \hat{c} \right) - \hat{c}\vec{\nabla}_s\cdot \vec{u}_s - \hat{c} \kappa \vec{u}\cdot\vec{n},
\label{equ:sharp_interface_model_1}
\end{equation}
where $\hat{c}$ is the interfacial scalar concentration (amount of scalar per unit area of the interface), $\vec{\nabla}_s=(I-\vec{n}\vec{n})\vec{\nabla}$ is the surface gradient, $\vec{n}$ is the interface normal, $\kappa$ is the curvature, $\vec{u}_s=(I-\vec{n}\vec{n})\vec{u}$ is the surface velocity, and $D$ is the diffusion coefficient. The evolution equation in Eq. \eqref{equ:sharp_interface_model_1} can also be rewritten in an equivalent form as \citep{teigen2009diffuse}
\begin{equation}
\frac{\partial \hat{c}}{\partial t} + \vec{\nabla}_s\cdot(\vec{u} \hat{c} ) = \vec{\nabla}_s \cdot \left(D \vec{\nabla}_s \hat{c} \right).
\label{equ:sharp_interface_model_2}
\end{equation}



\section{Phase-field model}


 
In this work, we use the recently developed accurate conservative phase-field/diffuse-interface model (ACDI) by \cite{jain2022accurate}, which is an Allen-Cahn-based second-order phase-field model given by
\begin{equation}
\frac{\partial \phi}{\partial t} + \vec{\nabla}\cdot(\vec{u}\phi) = \vec{\nabla}\cdot\left\{\Gamma\left\{\epsilon\vec{\nabla}\phi - \frac{1}{4} \left[1 - \tanh^2{\left(\frac{\psi}{2\epsilon}\right)}\right]\frac{\vec{\nabla} \psi}{|\vec{\nabla} \psi|}\right\}\right\},
\label{eq:ACDI}
\end{equation} 
where $\phi$ is the phase-field variable that represents the volume fraction, $\vec{u}$ is the velocity, $\Gamma$ represents the velocity-scale parameter, $\epsilon$ is the interface thickness scale parameter, and $\psi$ is an auxiliary signed-distance-like variable given by 
\begin{equation}
    \psi = \epsilon \ln\left(\frac{\phi + \varepsilon}{1 - \phi + \varepsilon}\right),
    \label{eq:psi}
\end{equation} 
where $\varepsilon=10^{-100}$ is a small number. The parameters are chosen to be $\Gamma \ge |\vec{u}|_{max}\ \mathrm{and}\ \epsilon > 0.5 \Delta x$, along with $\Delta t$ that satisfies the explicit Courant-Friedrich-Lewy criterion, to maintain the boundedness of $\phi$ \citep{jain2022accurate}. Here, $|\vec{u}|_{max}$ represents maximum value of the velocity in the domain and $\Delta x$ represents grid size.

The ACDI model has been recently gaining popularity \citep{hwang2023robust,brown2023phase,scapin2022,collis2022,liang2023lattice} due to its higher accuracy at lower cost compared to other phase-field models. This is because it maintains a sharper interface (with only one-to-two grid points across the interface) while being robust and conservative, without the need for any geometric treatment.  

 

\section{Proposed model for the transport of scalars/surfactants on evolving interfaces}

The proposed model for the transport of interface-confined scalars and insoluble surfactants on an evolving interface is 
\begin{equation}
\frac{\partial c}{\partial t} + \vec{\nabla}\cdot(\vec{u} c ) = \vec{\nabla} \cdot \left[D \left\{\vec{\nabla}c - \frac{2(0.5 - \phi) \vec{n} c}{\epsilon} \right\}\right],
\label{equ:interface_scalar_model}
\end{equation}
where $D$ is the diffusivity of the scalar, and $\vec{n}=\vec{\nabla}\phi/|\vec{\nabla}\phi|=\vec{\nabla}\psi/|\vec{\nabla}\psi|$ is the interface normal vector. The second term on the right-hand side (RHS) of Eq. \eqref{equ:interface_scalar_model} is an artificial sharpening term. The effect of this sharpening flux is to prevent the diffusion of the scalar on both sides of the interface and to confine it to the interface region, which is illustrated in Figure \ref{fig:sharp_flux}. 

Note the similarity of this sharpening flux to the model for transport of scalars in the bulk in Eq. \eqref{equ:bulk_scalar}, where the scalar is confined to one of the phases. The difference between the proposed model in Eq. \eqref{equ:interface_scalar_model} and the one in Eq. \eqref{equ:bulk_scalar} is the sharpening flux. The sharpening flux in Eq. \eqref{equ:bulk_scalar} acts along one direction and prevents the leakage of the scalar from one of the phases into the other phase, whereas in Eq. \eqref{equ:interface_scalar_model}, the sharpening flux acts in both directions, preventing the scalar from diffusing away from the interface region into either of the phases on both sides of the interface.  

\begin{figure}
    \centering
    \includegraphics[width=0.5\textwidth]{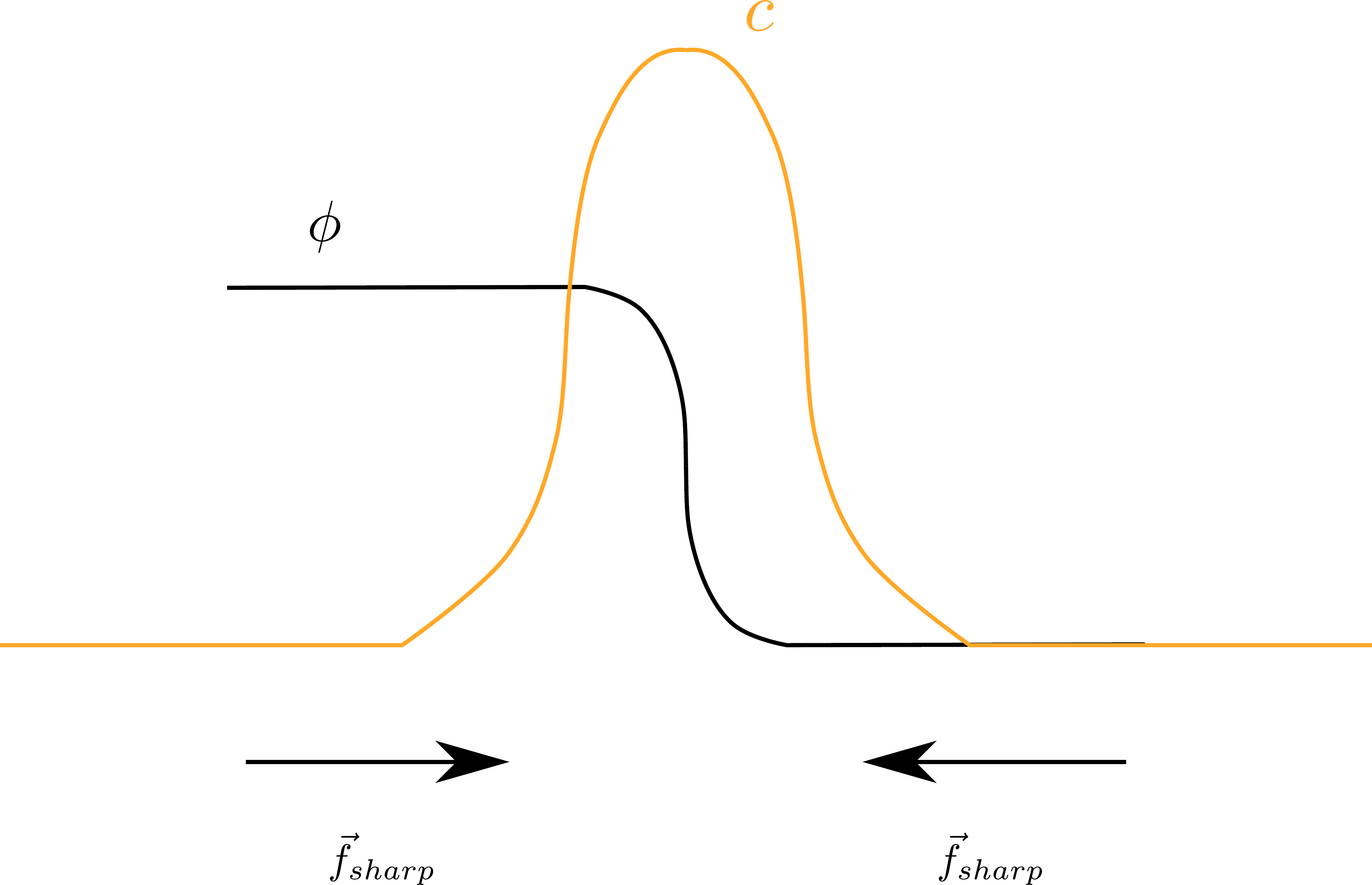}
    \caption{Schematic representing the effect of sharpening flux $\vec{f}_{sharp}={D 2(0.5 - \phi) \vec{n} c}/{\epsilon}$ in the model. Here, $\phi$ and $c$ are plotted at equilibrium to illustrate their equilibrium solutions: $\phi\sim \phi_{eq}$ and $c\sim \phi_{eq}'$.}
    \label{fig:sharp_flux}
\end{figure}

The proposed model in Eq. \eqref{equ:interface_scalar_model}
 is generalized in Appendix A, where the sharpness of the confinement of the scalar can be controlled.



%

\subsection{Consistency and equilibrium solution \label{sec:consistency}}

It is well known that the equilibrium solution (when $\Gamma\rightarrow\infty$) for the phase-field model in Eq. \eqref{eq:ACDI} is a hyperbolic tangent function given by
\begin{equation}
\phi_{eq} = \frac{1}{2} \left[1 + \tanh{\left(\frac{x}{2\epsilon}\right)}\right] \sim \tanh{\left(\frac{x}{2\epsilon}\right)}.
\label{equ:phi_sol}
\end{equation}
Now, taking a derivative of the equilibrium solution, we obtain
\begin{equation}
\phi_{eq}' \sim \frac{1}{4 \epsilon \cosh^2{\left(\frac{x}{2\epsilon}\right)}}.
\label{equ:phi_derv}
\end{equation}
This function is analogous to a Dirac delta function, a derivative of a step function, for the hyperbolic tangent function. Hence, a consistent transport model for interface-confined scalars and insoluble surfactants should possess an equilibrium solution of the form in Eq. \eqref{equ:phi_derv}. Both $\phi_{eq}$ and $\phi_{eq}'$ are shown in Figure \ref{fig:sharp_flux}.

To verify the equilibrium solution for the proposed model in Eq. \eqref{equ:interface_scalar_model}, let's assume steady state, and $\vec{u}=0$. In one dimension, the proposed model reduces to the form
\begin{equation}
0 = \vec{\nabla} \cdot \left[D \left\{\vec{\nabla}c - \frac{2(0.5 - \phi) \vec{n} c}{\epsilon} \right\}\right] \Rightarrow \frac{d^2 c}{d x^2} - \frac{1}{\epsilon} \frac{d\left\{2(0.5-\phi)c\right\}}{d x} = 0
\label{equ:c_steady}
\end{equation}
for $\vec{n}=+1$. Assuming the interface is at the origin and is in equilibrium, then
\begin{equation}
\phi = \phi_{eq} = \frac{e^{(x/\epsilon)}}{1 + e^{(x/\epsilon)}} = \frac{1}{2} \left\{1 + \tanh{\left(\frac{x}{2\epsilon}\right)}\right\}.
\label{equ:phi_sol}
\end{equation}
Using Eq. \eqref{equ:phi_sol} and solving for $c$ by integrating the Eq. \eqref{equ:c_steady} and using the boundary conditions 
\begin{equation}
    c = \Bigg\{
    \begin{aligned}
        & 0 \hspace{10mm} x \rightarrow -\infty \\
        & c_0 \hspace{10mm} x = 0,
    \end{aligned} \hspace{0.5cm} \mathrm{and} \hspace{0.5cm} \frac{dc}{dx}\rightarrow 0 \hspace{0.25cm} \mathrm{for} \hspace{0.25cm} x \rightarrow -\infty,
\end{equation}
we obtain
\begin{equation}
c = \frac{c_0}{\cosh^2{\left(\frac{x}{2\epsilon}\right)}}.
\label{equ:c_sol}
\end{equation}
Therefore, the equilibrium kernel function for the proposed model in Eq. \eqref{equ:interface_scalar_model} is indeed $c_{eq} \sim \phi_{eq}'$ [Eq. \eqref{equ:phi_derv}]. Hence, the proposed interface-confined scalar model is consistent with the phase-field model (or any model that admits a hyperbolic tangent function as its equilibrium solution). This results in the transport of the scalar along the interface without any unphysical numerical leakage into either of the phases on the two sides of the interface.

\subsection{Relationship with the sharp-interface surfactant-transport models}

The sharp-interface model in Eq. \eqref{equ:sharp_interface_model_1} can be rewritten in a distribution form \citep{teigen2009diffuse} as
\begin{equation}
\frac{\partial \left(\hat{c} \delta_s\right)}{\partial t} + \vec{\nabla} \cdot \left(\vec{u} \hat{c} \delta_s\right) = \vec{\nabla} \cdot \left(D \delta_s \vec{\nabla} \hat{c} \right),
\label{equ:teigen_interface_model}
\end{equation}
where $\delta_s$ is a surface delta function, defined as
\begin{equation}
    \int_{\gamma} \hat{c}\ d\gamma = \int_{\Omega} \hat{c} \delta_s\ d\Omega,
\end{equation}
where $\gamma$ is the interface and $\Omega$ is the domain.
The model in Eq. \eqref{equ:teigen_interface_model} can be solved directly by assuming a form for the surface delta function, an approach that was taken by \cite{teigen2009diffuse}. They used $\delta_s=3\sqrt{2}\phi^2(1 - \phi)^2/\epsilon$ and solved Eq. \eqref{equ:teigen_interface_model} with a Cahn-Hilliard phase-field model. However, this approach requires dividing $\hat{c} \delta_s$ by $\delta_s$ to compute $\hat{c}$ in the diffusion term in Eq. \eqref{equ:teigen_interface_model}, which could result in robustness issues.

The proposed model in Eq. \eqref{equ:interface_scalar_model} can also be derived starting from the transport equation in Eq. \eqref{equ:teigen_interface_model} by relating $c$ and $\hat{c}$ as
\begin{equation}
    \hat{c} = \frac{c}{\delta_s}.
\end{equation}
Using this relation in Eq. \eqref{equ:teigen_interface_model},
\begin{equation}
\frac{\partial c}{\partial t} + \vec{\nabla}\cdot(\vec{u} c ) = \vec{\nabla} \cdot \left[D \left\{\vec{\nabla}c - \frac{c}{\delta_s}\vec{\nabla} \delta_s \right\}\right],
\end{equation}
invoking the equilibrium interface condition
\begin{equation}
    \vec{\nabla}\phi = \frac{\phi (1 - \phi)}{\epsilon} \vec{n},
\end{equation}
and assuming the form for the surface delta function to be
\begin{equation}
    \delta_s = |\vec{\nabla}\phi|,
\end{equation}
we arrive at
\begin{equation}
\frac{\partial c}{\partial t} + \vec{\nabla}\cdot(\vec{u} c ) = \vec{\nabla} \cdot \left[D \left\{\vec{\nabla}c - \frac{2(0.5 - \phi) c}{\epsilon} \frac{\vec{\nabla} \phi}{|\vec{\nabla} \phi|} \right\}\right],
\end{equation}
which is the proposed model in Eq. \eqref{equ:interface_scalar_model} if $\vec{n}=\vec{\nabla}\phi/|\vec{\nabla}\phi|$. Note that it is not required to perform division by $\delta_s$ anywhere in the proposed model, which makes the method robust. Moreover, the surface gradient operator $\Vec{\nabla}_s$ is not used in the proposed model. Modeling such gradients accurately and efficiently requires infrastructure that is not needed with the proposed model, making it robust, easy to implement, and easily applicable for unstructured grids.

\section{Positivity \label{sec:positivity_proof}}

Following the proof of positivity in \cite{jain2023scalar}, the interfacial scalar or surfactant concentration $c$ in Eq. \eqref{equ:interface_scalar_model} can be shown to remain positive, i.e., $c^k_i\ge 0$ $\forall k\in\mathds{Z}^+$, where $k$ is the time-step index and $i$ is the grid index, provided the constraints
\begin{equation}
    \Delta x \le \left(\frac{2 D}{|u|_{\mathrm{max}} + \frac{D}{\epsilon}}\right),
    \label{eq:crossover}
\end{equation}
and 
\begin{equation}
      \Delta t \le \frac{\Delta x ^2}{2N_dD}
    \label{eq:boundtime}
\end{equation}
are satisfied, where $\Delta x$ is the grid-cell size, $\Delta t$ is the time-step size, $|u|_{\mathrm{max}}$ is the maximum fluid velocity in the domain, and $N_d$ is the number of dimensions. Note that this also requires $\phi^k_i$ to be bounded between $0$ and $1$, $\forall k\in\mathds{Z}^+$ and $\forall i$, which is guaranteed to be satisfied with the ACDI method \citep{jain2022accurate}.

If $\epsilon=\Delta x$, then the constraint in Eq. \eqref{eq:crossover} reduces to
\begin{equation}
\Delta x\le \frac{D}{|u|_{\mathrm{max}}}\ \text{or}\ Pe_c\le1,
\label{eq:positive}
\end{equation}
where $Pe_c=\Delta x |u|_{\mathrm{max}}/D$ is the cell-Peclet number. Similarly, for $\epsilon=0.75\Delta x$ the constraint is $Pe_c\le0.67$, and for $\epsilon=0.6\Delta x$ the constraint is $Pe_c\le0.33$.

\section{Numerical methods}

In this work, we use a second-order central scheme for spatial discretization and a fourth-order Runge-Kutta scheme for time stepping for the proposed model in Eq. \eqref{equ:interface_scalar_model}. A skew-symmetric-like flux-splitting approach \citep{jain2022kinetic} is adopted for the discretization of the ACDI method in Eq. \eqref{eq:ACDI}. 


\section{Simulation results}

In this section, simulations of the proposed model coupled with the ACDI method are presented.  
The simulations can be subdivided into three categories: (a) verification of the confinement of the scalar to the interface region in Section \ref{sec:confine}, (b) verification of the positivity of the scalar in Section \ref{sec:positive}, and (c) multidimensional simulations, which include a moving drop with an initially uniform scalar distribution (an extension of the one-dimensional cases) as well as a stationary drop with the diffusion of an initially nonuniform scalar distribution along the interface in Section \ref{sec:multi_d}.

\subsection{One-dimensional simulations}

In this section, one-dimensional simulations are presented, which act as verification of the proposed model. In all the simulations, a unit domain length of $L=1$ is used with a grid size of $\Delta x=0.01$, unless specified otherwise. A drop of radius $R=0.2$ is initially placed in the domain centered at $x_c=0.5$. The initial condition for the drop is given by $\phi_i = 0.5\left[1 - \tanh{\left\{\left(|x - 0.5| - 0.2\right)/(2\epsilon)\right\}}\right]$, where the subscript $i$ denotes $t=0$. 

\subsubsection{Confinement verification \label{sec:confine}}

To verify the effectiveness of the artificial sharpening term in the proposed model in Eq. \eqref{equ:interface_scalar_model} to confine the scalar to the interface region, we initialize the scalar uniformly within the drop in this section with a concentration of unity ($c_i=\phi_i$). Since the scalar is not permitted to dissolve into the bulk phase, we expect it to get reorganized and move to the interface region.

For the simulations in this section, a uniform velocity of $\vec{u}=0.5$ is prescribed, and both the drop and the scalar are advecting with this velocity field. The drop and the scalar are advected with a nonzero velocity field to verify the ability of the proposed model to reorganize the scalar field relative to the background flow field. The diffusivity is chosen to be $D=0.01$, so the $Pe_c=0.5$. 

\begin{figure}
    \centering
    \includegraphics[width=\textwidth]{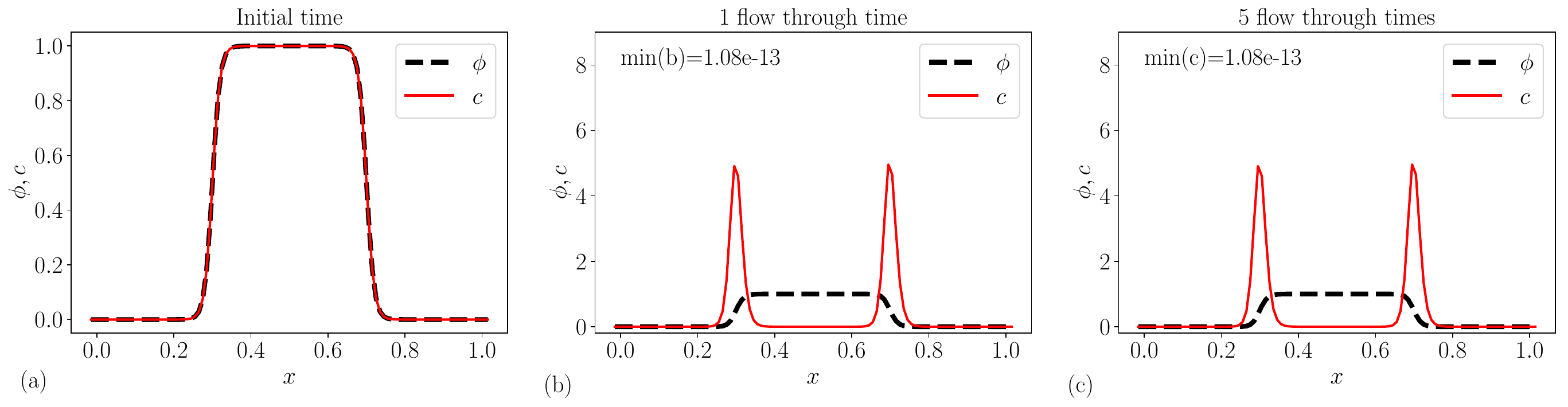}
    \caption{The advection of a drop along with an initially uniformly distributed scalar quantity (dissolved scalar in the bulk) inside the drop. (a) The initial drop and scalar setup, (b) the drop and scalar after 1 flow-through time at $t=2$, and (c) the drop and scalar after 5 flow-through times at $t=10$.}
    \label{fig:confine}
\end{figure}

Figure \ref{fig:confine} shows the evolution of the scalar with time. After 1 flow-through time, the scalar has reorganized to the interface region and has reached a steady state. After 5 flow-through times, the scalar is still confined to the interface region, as both the scalar and the drop are advecting with a velocity of $\vec{u}=0.5$. 

The artificial sharpening flux in Eq. \eqref{equ:interface_scalar_model} is responsible for the reorganization of the scalar. Therefore, the velocity associated with this reorganization of the scalar is given by $
\vec{u}_{re\_org}\approx D 2 (0.5-\phi) \vec{n}/\epsilon$. For the parameters chosen in this section, $\vec{u}_{re\_org} \approx 1$. Therefore, $\Vec{u}_{re\_org}$ is larger than the advection velocity $\Vec{u}$, and this is the case as long as the positivity criterion in Eq. \eqref{eq:positive} is satisfied. Hence, the sharpening flux will always dominate over any other background flow, thus resulting in the confinement of the scalar to the interface region.

\subsubsection{Positivity verification\label{sec:positive}}

In this section, the robustness of the positivity criterion in Eqs. \eqref{eq:crossover} and \eqref{eq:positive} is evaluated. The setup used here is the same as the one in Section \ref{sec:confine}. But two different diffusivities are chosen, $D=0.01$ and $D=0.0025$, which will result in $Pe_c=0.5$ and $Pe_c=2$, respectively. Since the simulation with $D=0.0025$ does not satisfy the positivity criterion, we expect the scalar to violate the positivity.

\begin{figure}
    \centering
    \includegraphics[width=\textwidth]{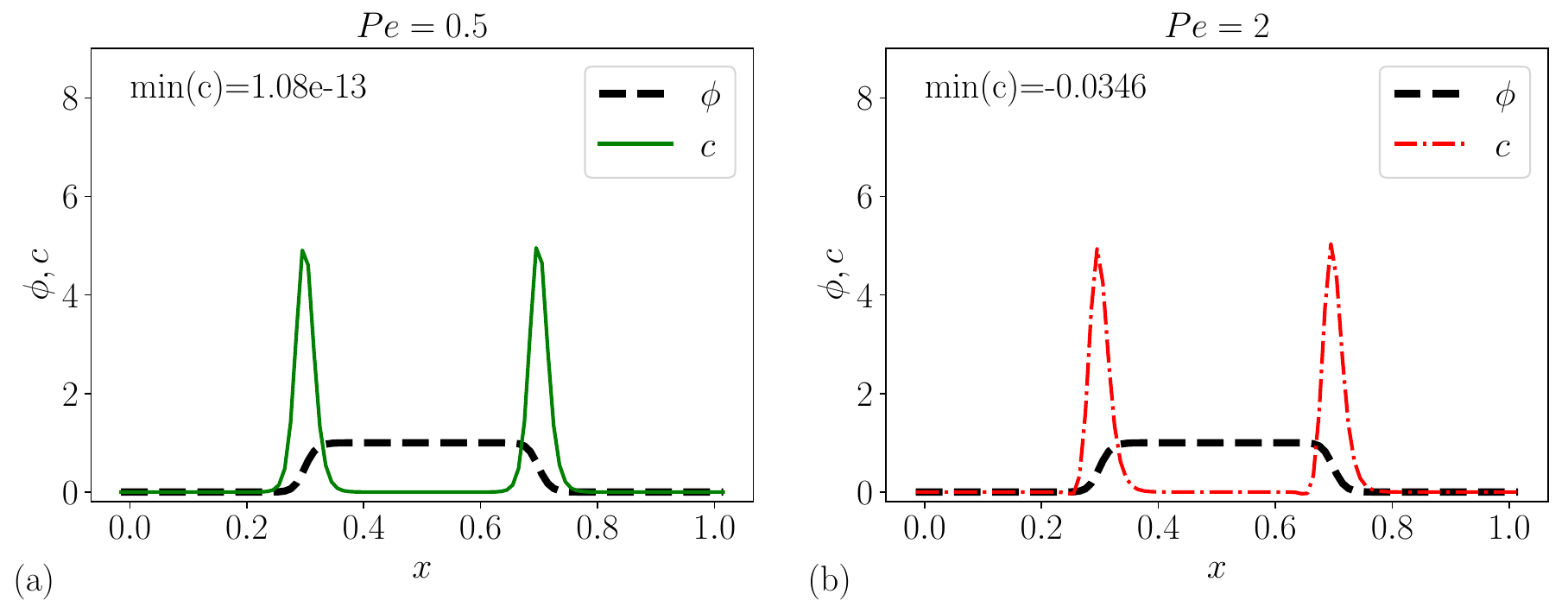}
    \caption{Final state of the drop and the scalar concentration field at time $t=2$. The two plots represent the two diffusivities chosen to test the positivity of the scalar: (a) $Pe_c=0.5$ and (b) $Pe_c=2$.}
    \label{fig:positive}
\end{figure}

Figure \ref{fig:positive} shows the final state of the drop and the scalar after 1 flow-through time. The minimum value of the scalar concentration field seen is also reported in the plots. As expected, the simulation with $Pe_c=2$ violates the positivity criterion, and therefore, negative values of the scalar concentration are observed.  


\subsection{Multidimensional simulations \label{sec:multi_d}}

In this section, the applicability of the proposed model for simulating multidimensional problems is tested. Two simulation setups in the subsequent sections are chosen: (a) Advecting drop\textemdash a two-dimensional version of the simulations in Section \ref{sec:confine}, and (b) surface diffusion of the scalar\textemdash a verification case, where the relative diffusion of the scalar along the interface is tested and compared against the analytical solutions.

\subsubsection{Advecting drop}

In this simulation, a unit square domain of size $L=1\times 1$ is used with a grid size of $\Delta x=0.01$. A drop of radius $R=0.2$ is initially placed in the domain center $(0.5,0.5)$. The initial condition for the drop is given by $\phi_i = 0.5\left[1 - \tanh{\left\{\left((x - 0.5)^2 + (y - 0.5)^2 - 0.2\right)/(2\epsilon)\right\}}\right]$. The scalar is initialized uniformly within the drop ($c_i=\phi_i$). A uniform velocity of $\Vec{u}=0.5$ is prescribed, and $Pe_c=0.5$. As was seen in Section \ref{sec:confine} for the one-dimensional setup, we expect the scalar to reorganize and move to the interface region.  

\begin{figure}
    \centering
    \includegraphics[width=\textwidth]{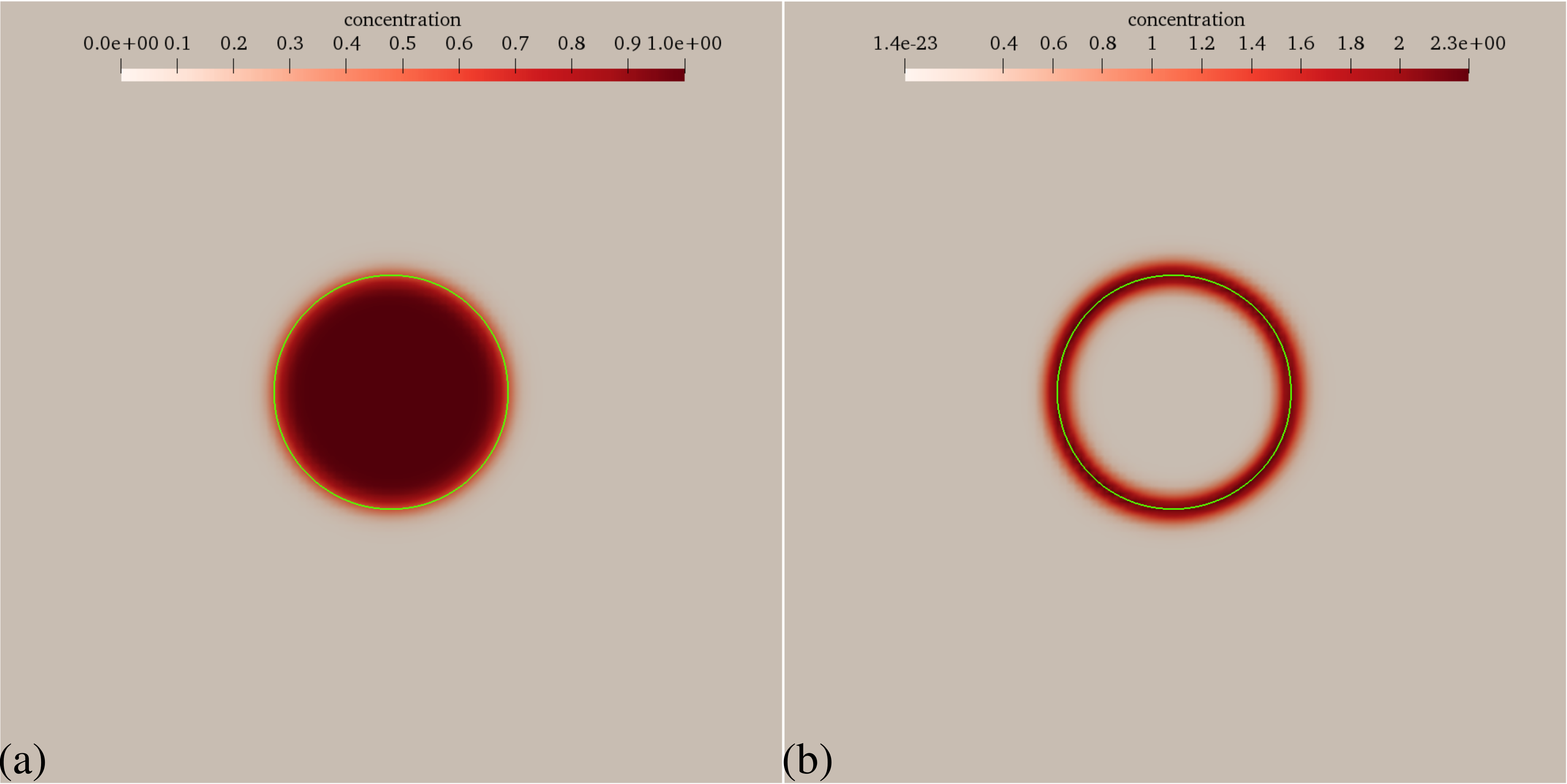}
    \caption{The advection of a two-dimensional drop along with an initially uniformly distributed scalar quantity (dissolved scalar in the bulk) inside the drop. (a) The drop and scalar setup at initial time. (b) The drop and scalar after 1 flow-through time at $t=2$. The solid green line is the isocontour of $\phi=0.5$ which represents the interface.}
    \label{fig:2d_confine}
\end{figure}

Figure \ref{fig:2d_confine} shows the scalar concentration and the drop at the initial and final time of $t=2$. As expected, the scalar reorganizes and moves to the interface region, since it is not allowed to dissolve in the bulk phase. This verifies the applicability of the proposed model in Eq. \eqref{equ:interface_scalar_model} in multidimensional problems without difficulty. 

\subsubsection{Surface diffusion of scalar}

In this section, the scalar is initially confined to the interface of a stationary drop, but with a non-uniform concentration profile along the interface. The accuracy of the proposed model to capture the diffusion of the scalar along the interface, while still maintaining the interfacial confinement, is verified by comparing against analytical solutions. 

The initial interfacial concentration (concentration per unit area of the interface) of the scalar is chosen to be 
\begin{equation}
    \hat{c}(\theta) = \frac{1}{2}\{1 - \cos\theta\}. 
\end{equation}
By solving a surface concentration equation in polar coordinates \citep{teigen2009diffuse}, the analytical solution for the diffusion of the scalar along the circular interface can be derived as
\begin{equation}
   \hat{c}(\theta,t) = \frac{1}{2} \left( 1 - e^{-\frac{D}{R^2}t} \cos \theta \right).
   \label{eq:surface_diff_anal}
\end{equation}

\begin{figure}
    \centering
    \includegraphics[width=0.8\textwidth]{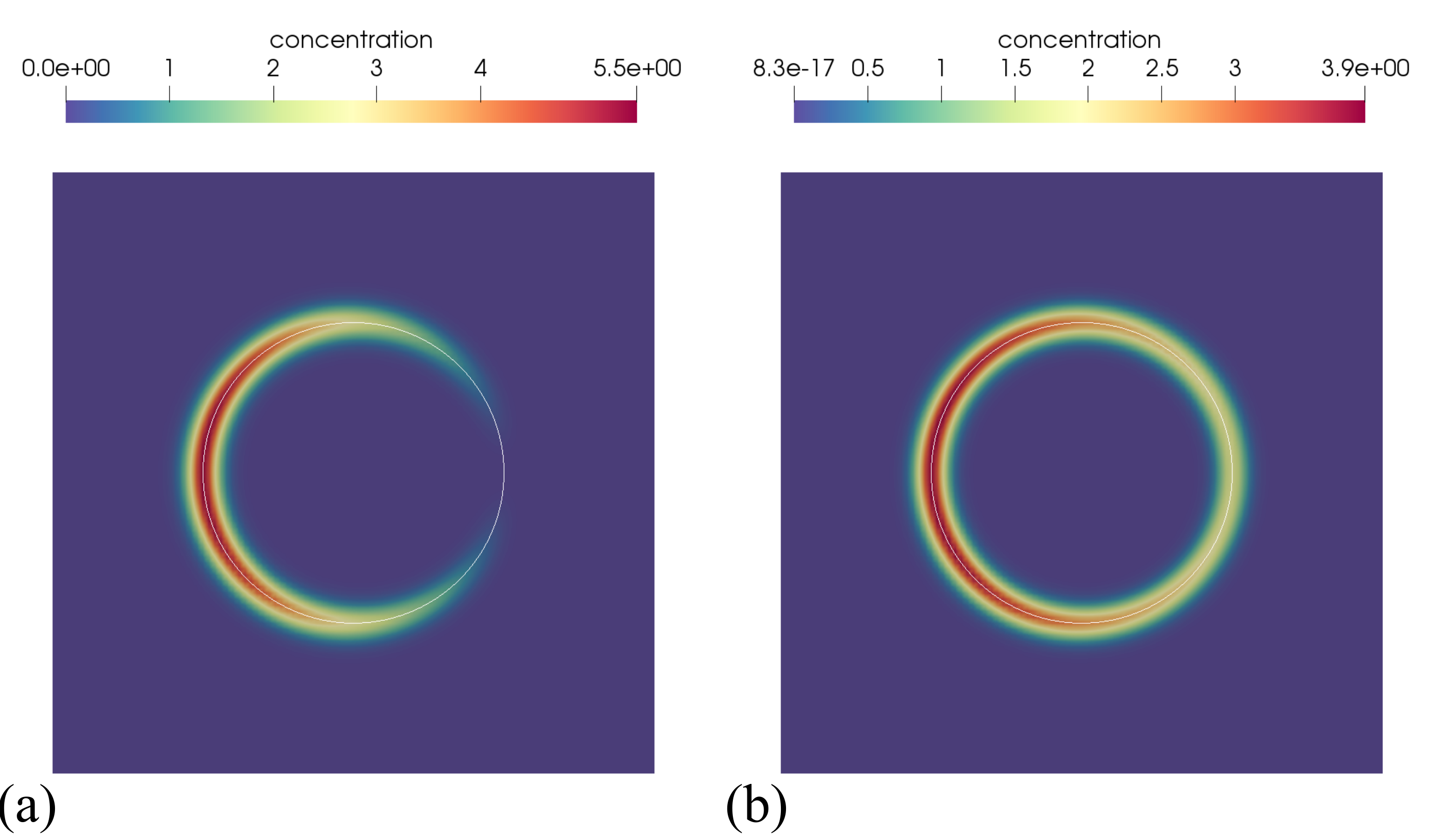}
    \caption{The surface diffusion of scalar on a two-dimensional stationary drop. (a) The drop and scalar setup at initial time. (b) The drop and scalar configuration at $t=1$.}
    \label{fig:2d_surface_diff}
\end{figure}

\begin{figure}
    \centering
    \includegraphics[width=0.8\textwidth]{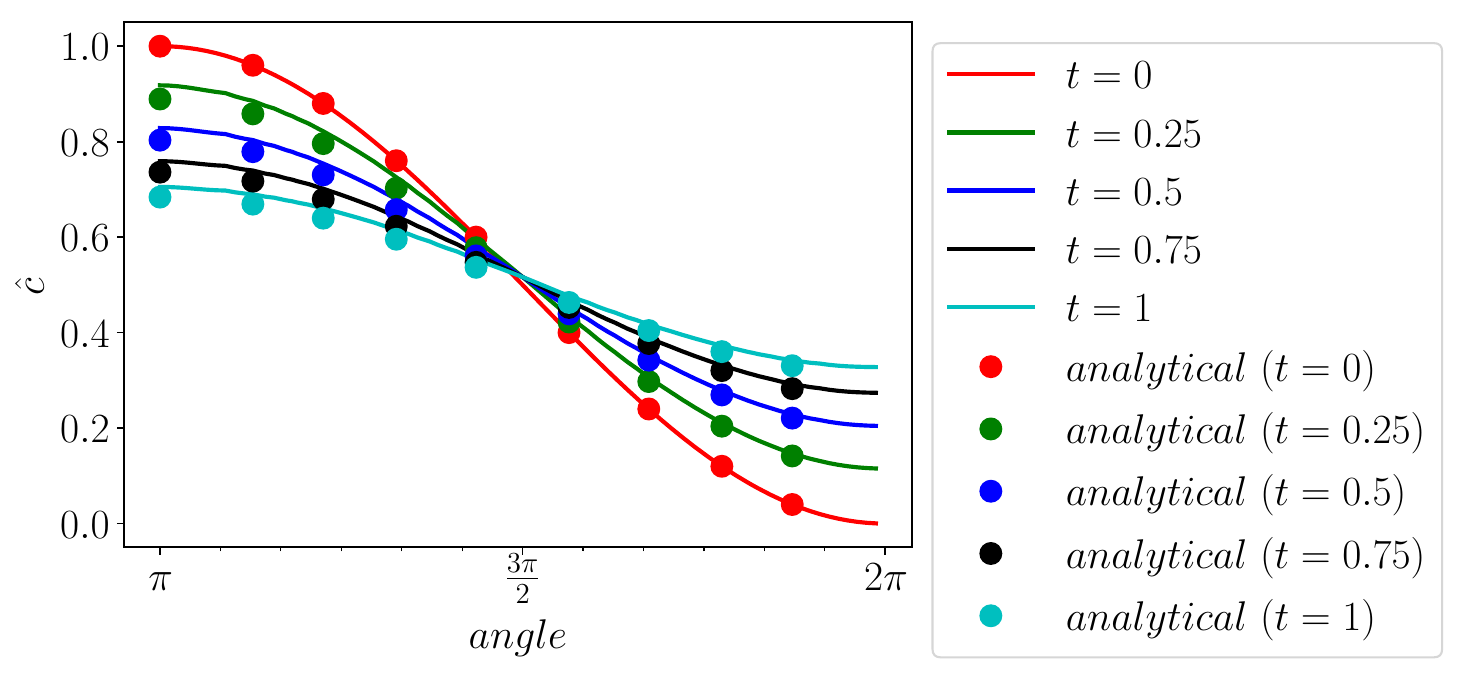}
    \caption{The local interfacial concentration of the scalar $\hat{c}$ along the drop at various time instances, computed using the proposed method, and its comparison with the analytical solution in Eq. \eqref{eq:surface_diff_anal}.}
    \label{fig:surface_diff}
\end{figure}

The simulation domain is chosen to be $[-2,2]\times[-2,2]$, with a grid size of $100\times100$, and the drop radius is $R=1$. Figure \ref{fig:2d_surface_diff} shows the scalar concentration at the initial and final time of $t=1$, which illustrates the diffusion of the scalar along the interface without any artificial leakage into the bulk phases. A quantitative comparison of the scalar concentration is shown in Figure \ref{fig:surface_diff} at various times, verifying the proposed model's capability to accurately simulate the transport of scalars that are confined to evolving material interfaces. 

\section{Conclusion}

In this work, a model for the transport of scalars confined to evolving material interfaces and insoluble surfactants in two-phase flows is developed. This model is solved with a second-order phase-field model; however, it can also be used with other interface-capturing methods. 
The scalar is shown to be consistently transported with the phase-field variable, resulting in a method that does not allow artificial leakage of the scalar into the bulk phases on either side of the interface. 
The model also results in positive scalar concentrations, a physical-realizability (robustness) condition, provided the given positivity criterion is satisfied. 

The proposed model was used to simulate transport of scalars that are confined to interfaces in a wide range of one-dimensional and multidimensional settings. The model was verified in terms of its capability to enforce confinement of the scalar to the interface region, the positivity of the scalar concentration, and its applicability for multidimensional problems. The accuracy of the model was also verified by comparing against analytical solutions.

\section*{Acknowledgments} 

S. S. J. acknowledges financial support from Boeing Co. 
S. S. J. is thankful to Ahmed Elnahhas, Makrand Khanwale, Ali Mani, and Parviz Moin for discussions.

\section*{Appendix A: Generalized model \label{sec:gen_model}}

The proposed model in Eq. \eqref{equ:interface_scalar_model} can be generalized as
\begin{equation}
\frac{\partial c}{\partial t} + \vec{\nabla}\cdot(\vec{u} c ) = \vec{\nabla} \cdot \left[D \left\{\vec{\nabla}c - \frac{a(0.5 - \phi) \vec{n} c}{\epsilon} \right\}\right],
\label{equ:gen_model}
\end{equation}
where $a$ is a constant. Theoretically, any value for $a$ is valid, which would still result in confinement of the scalar. But the choice of value of $a$ has consequences on the positivity of the scalar. The proof of positivity in Eq. \eqref{eq:crossover} in Section \ref{sec:positivity_proof} can be generalized to the model in Eq. \eqref{equ:gen_model} as
\begin{equation}
    \Delta x \le \left(\frac{2 D}{|u|_{\mathrm{max}} + \frac{D a}{2\epsilon}}\right).
    \label{eq:crossover_gen}
\end{equation}
If $\epsilon=\Delta x$, then the positivity constraint in Eq. \eqref{eq:crossover_gen} reduces to $\Delta x\le D (2 - a/2)/|u|_{\mathrm{max}}$. Hence, positivity can be achieved as long as $a<4$.
A higher value for $a$ results in sharper representation of the scalar at the interface. Hence, there is an upper limit on the value of $a$ beyond which the positivity (and robustness) is not guaranteed. 

\begin{figure}
    \centering
    \includegraphics[width=0.5\textwidth]{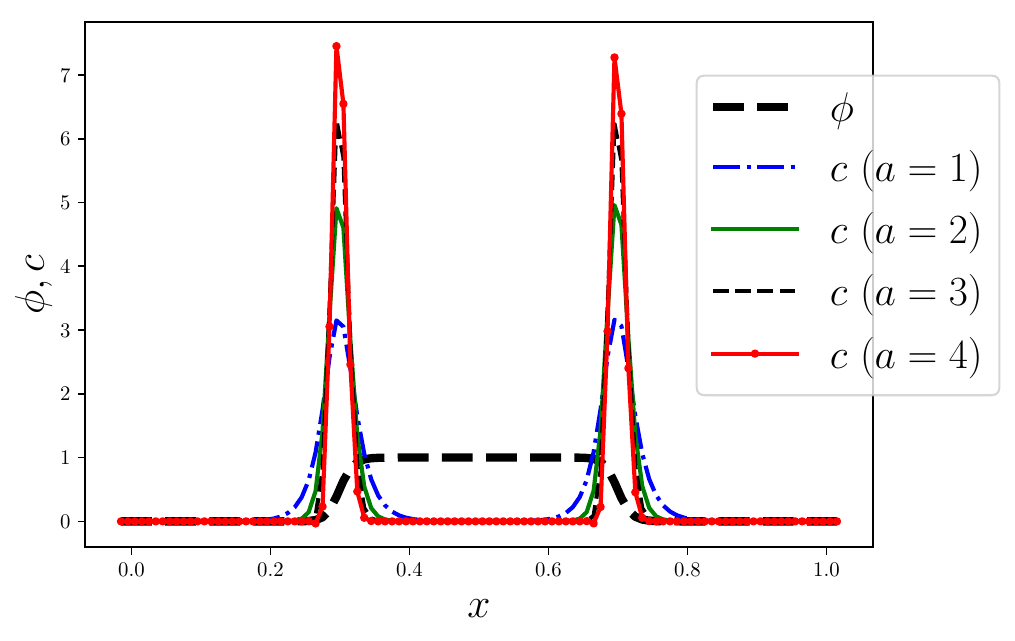}
    \caption{Final state of the drop and the scalar concentration field at time $t=2$ for various values of $a$ in Eq. \eqref{equ:gen_model}.}
    \label{fig:a_effect}
\end{figure}

The effect of using a value other than $2$ for $a$ in the generalized model in Eq. \eqref{equ:gen_model} is illustrated below. The setup is the same as the one in Section \ref{sec:confine} with $Pe_c=0.5$, and four different values $a=1,2,3,$ and $4$ are tested. Figure \ref{fig:a_effect} shows the final state of the drop and the scalar after 1 flow-through time. 
The positivity of the scalar was verified for the cases $a=1,2$, and $3$, and the positivity was violated for $a=4$ as expected.

With an increase in $a$, the scalar is more concentrated at the interface. One might think that this could result in an improved accuracy due to a sharper representation of the interface-confined scalar concentration. However, note that the equilibrium solution (following the procedure in Section \ref{sec:consistency}) for the model in Eq. \eqref{equ:gen_model} is
\begin{equation}
c = \frac{c_0}{\cosh^a{\left(\frac{x}{2\epsilon}\right)}},
\label{equ:c_sol_gen}
\end{equation} 
which is only consistent with the equilibrium solution of the phase-field model [Eq. \eqref{equ:phi_derv}] if $a=2$.
Hence, only the value of $a=2$ will result in an interface-confined scalar model that is exactly consistent with the phase-field model, as described in Section \ref{sec:gen_model}, and therefore $a=2$ is the recommended value. 

\bibliographystyle{model1-num-names}
\bibliography{two_phase}

\begin{thebibliography}{61}
\expandafter\ifx\csname natexlab\endcsname\relax\def\natexlab#1{#1}\fi
\providecommand{\bibinfo}[2]{#2}
\ifx\xfnm\relax \def\xfnm[#1]{\unskip,\space#1}\fi
\bibitem[{Jain(2022)}]{jain2022accurate}
\bibinfo{author}{S.~S. Jain},
\newblock \bibinfo{title}{Accurate conservative phase-field method for
  simulation of two-phase flows},
\newblock \bibinfo{journal}{J. Comput. Phys.} \bibinfo{volume}{469}
  (\bibinfo{year}{2022}) \bibinfo{pages}{111529}.
\bibitem[{Chu and Bazant(2007)}]{chu2007surface}
\bibinfo{author}{K.~T. Chu}, \bibinfo{author}{M.~Z. Bazant},
\newblock \bibinfo{title}{Surface conservation laws at microscopically diffuse
  interfaces},
\newblock \bibinfo{journal}{Journal of colloid and interface science}
  \bibinfo{volume}{315} (\bibinfo{year}{2007}) \bibinfo{pages}{319--329}.
\bibitem[{Hargreaves(2007)}]{hargreaves2007chemical}
\bibinfo{author}{A.~E. Hargreaves},
\newblock \bibinfo{title}{Chemical formulation: an overview of surfactant based
  chemical preparations used in everyday life}  (\bibinfo{year}{2007}).
\bibitem[{Defay et~al.(1966)Defay, Prigogine, and Bellemans}]{defay1966surface}
\bibinfo{author}{R.~Defay}, \bibinfo{author}{I.~Prigogine},
  \bibinfo{author}{A.~Bellemans}, \bibinfo{title}{Surface tension and
  adsorption}, \bibinfo{publisher}{Wiley}, \bibinfo{year}{1966}.
\bibitem[{Eggleton et~al.(2001)Eggleton, Tsai, and Stebe}]{eggleton2001tip}
\bibinfo{author}{C.~D. Eggleton}, \bibinfo{author}{T.-M. Tsai},
  \bibinfo{author}{K.~J. Stebe},
\newblock \bibinfo{title}{Tip streaming from a drop in the presence of
  surfactants},
\newblock \bibinfo{journal}{Physical review letters} \bibinfo{volume}{87}
  (\bibinfo{year}{2001}) \bibinfo{pages}{048302}.
\bibitem[{Booty and Siegel(2005)}]{booty2005steady}
\bibinfo{author}{M.~Booty}, \bibinfo{author}{M.~Siegel},
\newblock \bibinfo{title}{Steady deformation and tip-streaming of a slender
  bubble with surfactant in an extensional flow},
\newblock \bibinfo{journal}{Journal of Fluid Mechanics} \bibinfo{volume}{544}
  (\bibinfo{year}{2005}) \bibinfo{pages}{243--275}.
\bibitem[{Baret(2012)}]{baret2012surfactants}
\bibinfo{author}{J.-C. Baret},
\newblock \bibinfo{title}{Surfactants in droplet-based microfluidics},
\newblock \bibinfo{journal}{Lab on a Chip} \bibinfo{volume}{12}
  (\bibinfo{year}{2012}) \bibinfo{pages}{422--433}.
\bibitem[{Pit et~al.(2015)Pit, Duits, and Mugele}]{pit2015droplet}
\bibinfo{author}{A.~M. Pit}, \bibinfo{author}{M.~H. Duits},
  \bibinfo{author}{F.~Mugele},
\newblock \bibinfo{title}{Droplet manipulations in two phase flow
  microfluidics},
\newblock \bibinfo{journal}{Micromachines} \bibinfo{volume}{6}
  (\bibinfo{year}{2015}) \bibinfo{pages}{1768--1793}.
\bibitem[{Manfield et~al.(1999)Manfield, Lawrence, and
  Hewitt}]{manfield1999drag}
\bibinfo{author}{P.~Manfield}, \bibinfo{author}{C.~Lawrence},
  \bibinfo{author}{G.~F. Hewitt},
\newblock \bibinfo{title}{Drag reduction with additives in multiphase flow: a
  literature survey},
\newblock \bibinfo{journal}{multiphase Science and Technology}
  \bibinfo{volume}{11} (\bibinfo{year}{1999}).
\bibitem[{Yap and Gaver~III(1998)}]{yap1998influence}
\bibinfo{author}{D.~Y. Yap}, \bibinfo{author}{D.~P. Gaver~III},
\newblock \bibinfo{title}{The influence of surfactant on two-phase flow in a
  flexible-walled channel under bulk equilibrium conditions},
\newblock \bibinfo{journal}{Physics of Fluids} \bibinfo{volume}{10}
  (\bibinfo{year}{1998}) \bibinfo{pages}{1846--1863}.
\bibitem[{Stone(1990)}]{stone1990simple}
\bibinfo{author}{H.~Stone},
\newblock \bibinfo{title}{A simple derivation of the time-dependent
  convective-diffusion equation for surfactant transport along a deforming
  interface},
\newblock \bibinfo{journal}{Physics of Fluids A: Fluid Dynamics}
  \bibinfo{volume}{2} (\bibinfo{year}{1990}) \bibinfo{pages}{111--112}.
\bibitem[{Wong et~al.(1996)Wong, Rumschitzki, and
  Maldarelli}]{wong1996surfactant}
\bibinfo{author}{H.~Wong}, \bibinfo{author}{D.~Rumschitzki},
  \bibinfo{author}{C.~Maldarelli},
\newblock \bibinfo{title}{On the surfactant mass balance at a deforming fluid
  interface},
\newblock \bibinfo{journal}{Physics of Fluids} \bibinfo{volume}{8}
  (\bibinfo{year}{1996}) \bibinfo{pages}{3203--3204}.
\bibitem[{Stone and Leal(1990)}]{stone1990effects}
\bibinfo{author}{H.~A. Stone}, \bibinfo{author}{L.~G. Leal},
\newblock \bibinfo{title}{The effects of surfactants on drop deformation and
  breakup},
\newblock \bibinfo{journal}{Journal of Fluid Mechanics} \bibinfo{volume}{220}
  (\bibinfo{year}{1990}) \bibinfo{pages}{161--186}.
\bibitem[{Milliken et~al.(1993)Milliken, Stone, and Leal}]{milliken1993effect}
\bibinfo{author}{W.~Milliken}, \bibinfo{author}{H.~A. Stone},
  \bibinfo{author}{L.~Leal},
\newblock \bibinfo{title}{The effect of surfactant on the transient motion of
  newtonian drops},
\newblock \bibinfo{journal}{Physics of Fluids A: Fluid Dynamics}
  \bibinfo{volume}{5} (\bibinfo{year}{1993}) \bibinfo{pages}{69--79}.
\bibitem[{Milliken and Leal(1994)}]{milliken1994influence}
\bibinfo{author}{W.~J. Milliken}, \bibinfo{author}{L.~G. Leal},
\newblock \bibinfo{title}{The influence of surfactant on the deformation and
  breakup of a viscous drop: The effect of surfactant solubility},
\newblock \bibinfo{journal}{Journal of Colloid and Interface Science}
  \bibinfo{volume}{166} (\bibinfo{year}{1994}) \bibinfo{pages}{275--285}.
\bibitem[{Pawar and Stebe(1996)}]{pawar1996marangoni}
\bibinfo{author}{Y.~Pawar}, \bibinfo{author}{K.~J. Stebe},
\newblock \bibinfo{title}{Marangoni effects on drop deformation in an
  extensional flow: The role of surfactant physical chemistry. i. insoluble
  surfactants},
\newblock \bibinfo{journal}{Physics of Fluids} \bibinfo{volume}{8}
  (\bibinfo{year}{1996}) \bibinfo{pages}{1738--1751}.
\bibitem[{Siegel(1999)}]{siegel1999influence}
\bibinfo{author}{M.~Siegel},
\newblock \bibinfo{title}{Influence of surfactant on rounded and pointed
  bubbles in two-dimensional stokes flow},
\newblock \bibinfo{journal}{SIAM Journal on Applied Mathematics}
  \bibinfo{volume}{59} (\bibinfo{year}{1999}) \bibinfo{pages}{1998--2027}.
\bibitem[{Cuenot et~al.(1997)Cuenot, Magnaudet, and
  Spennato}]{cuenot1997effects}
\bibinfo{author}{B.~Cuenot}, \bibinfo{author}{J.~Magnaudet},
  \bibinfo{author}{B.~Spennato},
\newblock \bibinfo{title}{The effects of slightly soluble surfactants on the
  flow around a spherical bubble},
\newblock \bibinfo{journal}{Journal of fluid mechanics} \bibinfo{volume}{339}
  (\bibinfo{year}{1997}) \bibinfo{pages}{25--53}.
\bibitem[{Li and Pozrikidis(1997)}]{li1997effect}
\bibinfo{author}{X.~Li}, \bibinfo{author}{C.~Pozrikidis},
\newblock \bibinfo{title}{The effect of surfactants on drop deformation and on
  the rheology of dilute emulsions in stokes flow},
\newblock \bibinfo{journal}{Journal of fluid mechanics} \bibinfo{volume}{341}
  (\bibinfo{year}{1997}) \bibinfo{pages}{165--194}.
\bibitem[{Yon and Pozrikidis(1998)}]{yon1998finite}
\bibinfo{author}{S.~Yon}, \bibinfo{author}{C.~Pozrikidis},
\newblock \bibinfo{title}{A finite-volume/boundary-element method for flow past
  interfaces in the presence of surfactants, with application to shear flow
  past a viscous drop},
\newblock \bibinfo{journal}{Computers \& fluids} \bibinfo{volume}{27}
  (\bibinfo{year}{1998}) \bibinfo{pages}{879--902}.
\bibitem[{Eggleton et~al.(1999)Eggleton, Pawar, and
  Stebe}]{eggleton1999insoluble}
\bibinfo{author}{C.~D. Eggleton}, \bibinfo{author}{Y.~P. Pawar},
  \bibinfo{author}{K.~J. Stebe},
\newblock \bibinfo{title}{Insoluble surfactants on a drop in an extensional
  flow: a generalization of the stagnated surface limit to deforming
  interfaces},
\newblock \bibinfo{journal}{Journal of Fluid Mechanics} \bibinfo{volume}{385}
  (\bibinfo{year}{1999}) \bibinfo{pages}{79--99}.
\bibitem[{Hsu et~al.(2019)Hsu, Chu, Lai, and Tsai}]{hsu2019coupled}
\bibinfo{author}{S.-H. Hsu}, \bibinfo{author}{J.~Chu}, \bibinfo{author}{M.-C.
  Lai}, \bibinfo{author}{R.~Tsai},
\newblock \bibinfo{title}{A coupled grid based particle and implicit boundary
  integral method for two-phase flows with insoluble surfactant},
\newblock \bibinfo{journal}{Journal of Computational Physics}
  \bibinfo{volume}{395} (\bibinfo{year}{2019}) \bibinfo{pages}{747--764}.
\bibitem[{Renardy et~al.(2002)Renardy, Renardy, and Cristini}]{renardy2002new}
\bibinfo{author}{Y.~Y. Renardy}, \bibinfo{author}{M.~Renardy},
  \bibinfo{author}{V.~Cristini},
\newblock \bibinfo{title}{A new volume-of-fluid formulation for surfactants and
  simulations of drop deformation under shear at a low viscosity ratio},
\newblock \bibinfo{journal}{European Journal of Mechanics-B/Fluids}
  \bibinfo{volume}{21} (\bibinfo{year}{2002}) \bibinfo{pages}{49--59}.
\bibitem[{Drumright-Clarke and Renardy(2004)}]{drumright2004effect}
\bibinfo{author}{M.~Drumright-Clarke}, \bibinfo{author}{Y.~Renardy},
\newblock \bibinfo{title}{The effect of insoluble surfactant at dilute
  concentration on drop breakup under shear with inertia},
\newblock \bibinfo{journal}{Physics of fluids} \bibinfo{volume}{16}
  (\bibinfo{year}{2004}) \bibinfo{pages}{14--21}.
\bibitem[{James and Lowengrub(2004)}]{james2004surfactant}
\bibinfo{author}{A.~J. James}, \bibinfo{author}{J.~Lowengrub},
\newblock \bibinfo{title}{A surfactant-conserving volume-of-fluid method for
  interfacial flows with insoluble surfactant},
\newblock \bibinfo{journal}{Journal of computational physics}
  \bibinfo{volume}{201} (\bibinfo{year}{2004}) \bibinfo{pages}{685--722}.
\bibitem[{Pozrikidis(2004)}]{pozrikidis2004finite}
\bibinfo{author}{C.~Pozrikidis},
\newblock \bibinfo{title}{A finite-element method for interfacial surfactant
  transport, with application to the flow-induced deformation of a viscous
  drop},
\newblock \bibinfo{journal}{Journal of engineering mathematics}
  \bibinfo{volume}{49} (\bibinfo{year}{2004}) \bibinfo{pages}{163--180}.
\bibitem[{Ganesan and Tobiska(2009)}]{ganesan2009coupled}
\bibinfo{author}{S.~Ganesan}, \bibinfo{author}{L.~Tobiska},
\newblock \bibinfo{title}{A coupled arbitrary lagrangian--eulerian and
  lagrangian method for computation of free surface flows with insoluble
  surfactants},
\newblock \bibinfo{journal}{Journal of Computational Physics}
  \bibinfo{volume}{228} (\bibinfo{year}{2009}) \bibinfo{pages}{2859--2873}.
\bibitem[{Venkatesan et~al.(2019)Venkatesan, Padmanabhan, and
  Ganesan}]{venkatesan2019simulation}
\bibinfo{author}{J.~Venkatesan}, \bibinfo{author}{A.~Padmanabhan},
  \bibinfo{author}{S.~Ganesan},
\newblock \bibinfo{title}{Simulation of viscoelastic two-phase flows with
  insoluble surfactants},
\newblock \bibinfo{journal}{Journal of Non-Newtonian Fluid Mechanics}
  \bibinfo{volume}{267} (\bibinfo{year}{2019}) \bibinfo{pages}{61--77}.
\bibitem[{Frachon and Zahedi(2023)}]{frachon2023cut}
\bibinfo{author}{T.~Frachon}, \bibinfo{author}{S.~Zahedi},
\newblock \bibinfo{title}{A cut finite element method for two-phase flows with
  insoluble surfactants},
\newblock \bibinfo{journal}{Journal of Computational Physics}
  \bibinfo{volume}{473} (\bibinfo{year}{2023}) \bibinfo{pages}{111734}.
\bibitem[{Khatri and Tornberg(2011)}]{khatri2011numerical}
\bibinfo{author}{S.~Khatri}, \bibinfo{author}{A.-K. Tornberg},
\newblock \bibinfo{title}{A numerical method for two phase flows with insoluble
  surfactants},
\newblock \bibinfo{journal}{Computers \& fluids} \bibinfo{volume}{49}
  (\bibinfo{year}{2011}) \bibinfo{pages}{150--165}.
\bibitem[{Xu and Zhao(2003)}]{xu2003eulerian}
\bibinfo{author}{J.-J. Xu}, \bibinfo{author}{H.-K. Zhao},
\newblock \bibinfo{title}{An eulerian formulation for solving partial
  differential equations along a moving interface},
\newblock \bibinfo{journal}{Journal of Scientific Computing}
  \bibinfo{volume}{19} (\bibinfo{year}{2003}) \bibinfo{pages}{573--594}.
\bibitem[{Xu et~al.(2006)Xu, Li, Lowengrub, and Zhao}]{xu2006level}
\bibinfo{author}{J.-J. Xu}, \bibinfo{author}{Z.~Li},
  \bibinfo{author}{J.~Lowengrub}, \bibinfo{author}{H.~Zhao},
\newblock \bibinfo{title}{A level-set method for interfacial flows with
  surfactant},
\newblock \bibinfo{journal}{Journal of Computational Physics}
  \bibinfo{volume}{212} (\bibinfo{year}{2006}) \bibinfo{pages}{590--616}.
\bibitem[{Xu et~al.(2012)Xu, Yang, and Lowengrub}]{xu2012level}
\bibinfo{author}{J.-J. Xu}, \bibinfo{author}{Y.~Yang},
  \bibinfo{author}{J.~Lowengrub},
\newblock \bibinfo{title}{A level-set continuum method for two-phase flows with
  insoluble surfactant},
\newblock \bibinfo{journal}{Journal of Computational Physics}
  \bibinfo{volume}{231} (\bibinfo{year}{2012}) \bibinfo{pages}{5897--5909}.
\bibitem[{de~Jesus et~al.(2015)de~Jesus, Roma, Pivello, Villar, and
  da~Silveira-Neto}]{de20153d}
\bibinfo{author}{W.~C. de~Jesus}, \bibinfo{author}{A.~M. Roma},
  \bibinfo{author}{M.~R. Pivello}, \bibinfo{author}{M.~M. Villar},
  \bibinfo{author}{A.~da~Silveira-Neto},
\newblock \bibinfo{title}{A 3d front-tracking approach for simulation of a
  two-phase fluid with insoluble surfactant},
\newblock \bibinfo{journal}{Journal of Computational Physics}
  \bibinfo{volume}{281} (\bibinfo{year}{2015}) \bibinfo{pages}{403--420}.
\bibitem[{Lai et~al.(2008)Lai, Tseng, and Huang}]{lai2008immersed}
\bibinfo{author}{M.-C. Lai}, \bibinfo{author}{Y.-H. Tseng},
  \bibinfo{author}{H.~Huang},
\newblock \bibinfo{title}{An immersed boundary method for interfacial flows
  with insoluble surfactant},
\newblock \bibinfo{journal}{Journal of Computational Physics}
  \bibinfo{volume}{227} (\bibinfo{year}{2008}) \bibinfo{pages}{7279--7293}.
\bibitem[{Ceniceros(2003)}]{ceniceros2003effects}
\bibinfo{author}{H.~D. Ceniceros},
\newblock \bibinfo{title}{The effects of surfactants on the formation and
  evolution of capillary waves},
\newblock \bibinfo{journal}{Physics of Fluids} \bibinfo{volume}{15}
  (\bibinfo{year}{2003}) \bibinfo{pages}{245--256}.
\bibitem[{Cui(2011)}]{cui2011computational}
\bibinfo{author}{Y.~Cui}, \bibinfo{title}{A computational fluid dynamics study
  of two-phase flows in the presence of surfactants},
  \bibinfo{publisher}{University of New Hampshire}, \bibinfo{year}{2011}.
\bibitem[{Van~der Sman and Van~der Graaf(2006)}]{van2006diffuse}
\bibinfo{author}{R.~Van~der Sman}, \bibinfo{author}{S.~Van~der Graaf},
\newblock \bibinfo{title}{Diffuse interface model of surfactant adsorption onto
  flat and droplet interfaces},
\newblock \bibinfo{journal}{Rheologica acta} \bibinfo{volume}{46}
  (\bibinfo{year}{2006}) \bibinfo{pages}{3--11}.
\bibitem[{Yun et~al.(2014)Yun, Li, and Kim}]{yun2014new}
\bibinfo{author}{A.~Yun}, \bibinfo{author}{Y.~Li}, \bibinfo{author}{J.~Kim},
\newblock \bibinfo{title}{A new phase-field model for a water--oil-surfactant
  system},
\newblock \bibinfo{journal}{Applied Mathematics and Computation}
  \bibinfo{volume}{229} (\bibinfo{year}{2014}) \bibinfo{pages}{422--432}.
\bibitem[{Engblom et~al.(2013)Engblom, Do-Quang, Amberg, and
  Tornberg}]{engblom2013diffuse}
\bibinfo{author}{S.~Engblom}, \bibinfo{author}{M.~Do-Quang},
  \bibinfo{author}{G.~Amberg}, \bibinfo{author}{A.-K. Tornberg},
\newblock \bibinfo{title}{On diffuse interface modeling and simulation of
  surfactants in two-phase fluid flow},
\newblock \bibinfo{journal}{Communications in Computational Physics}
  \bibinfo{volume}{14} (\bibinfo{year}{2013}) \bibinfo{pages}{879--915}.
\bibitem[{Abels et~al.(2019)Abels, Garcke, and Weber}]{abels2019existence}
\bibinfo{author}{H.~Abels}, \bibinfo{author}{H.~Garcke},
  \bibinfo{author}{J.~Weber},
\newblock \bibinfo{title}{Existence of weak solutions for a diffuse interface
  model for two-phase flow with surfactants.},
\newblock \bibinfo{journal}{Communications on Pure \& Applied Analysis}
  \bibinfo{volume}{18} (\bibinfo{year}{2019}).
\bibitem[{Di~Primio et~al.(2022)Di~Primio, Grasselli, and Wu}]{di2022well}
\bibinfo{author}{A.~Di~Primio}, \bibinfo{author}{M.~Grasselli},
  \bibinfo{author}{H.~Wu},
\newblock \bibinfo{title}{Well-posedness for a navier-stokes-cahn-hilliard
  system for incompressible two-phase flows with surfactant},
\newblock \bibinfo{journal}{arXiv preprint arXiv:2201.09022}
  (\bibinfo{year}{2022}).
\bibitem[{Teigen et~al.(2009)Teigen, Li, Lowengrub, Wang, and
  Voigt}]{teigen2009diffuse}
\bibinfo{author}{K.~E. Teigen}, \bibinfo{author}{X.~Li},
  \bibinfo{author}{J.~Lowengrub}, \bibinfo{author}{F.~Wang},
  \bibinfo{author}{A.~Voigt},
\newblock \bibinfo{title}{A diffuse-interface approach for modeling transport,
  diffusion and adsorption/desorption of material quantities on a deformable
  interface},
\newblock \bibinfo{journal}{Communications in mathematical sciences}
  \bibinfo{volume}{4} (\bibinfo{year}{2009}) \bibinfo{pages}{1009}.
\bibitem[{Teigen et~al.(2011)Teigen, Song, Lowengrub, and
  Voigt}]{teigen2011diffuse}
\bibinfo{author}{K.~E. Teigen}, \bibinfo{author}{P.~Song},
  \bibinfo{author}{J.~Lowengrub}, \bibinfo{author}{A.~Voigt},
\newblock \bibinfo{title}{A diffuse-interface method for two-phase flows with
  soluble surfactants},
\newblock \bibinfo{journal}{Journal of computational physics}
  \bibinfo{volume}{230} (\bibinfo{year}{2011}) \bibinfo{pages}{375--393}.
\bibitem[{Garcke et~al.(2014)Garcke, Lam, and Stinner}]{garcke2014diffuse}
\bibinfo{author}{H.~Garcke}, \bibinfo{author}{K.~F. Lam},
  \bibinfo{author}{B.~Stinner},
\newblock \bibinfo{title}{Diffuse interface modelling of soluble surfactants in
  two-phase flow},
\newblock \bibinfo{journal}{Communications in Mathematical Sciences}
  \bibinfo{volume}{12} (\bibinfo{year}{2014}) \bibinfo{pages}{1475--1522}.
\bibitem[{Ray et~al.(2021)Ray, Liu, and Riviere}]{ray2021discontinuous}
\bibinfo{author}{D.~Ray}, \bibinfo{author}{C.~Liu},
  \bibinfo{author}{B.~Riviere},
\newblock \bibinfo{title}{A discontinuous galerkin method for a
  diffuse-interface model of immiscible two-phase flows with soluble
  surfactant},
\newblock \bibinfo{journal}{Computational Geosciences} \bibinfo{volume}{25}
  (\bibinfo{year}{2021}) \bibinfo{pages}{1775--1792}.
\bibitem[{Soligo et~al.(2019)Soligo, Roccon, and Soldati}]{soligo2019breakage}
\bibinfo{author}{G.~Soligo}, \bibinfo{author}{A.~Roccon},
  \bibinfo{author}{A.~Soldati},
\newblock \bibinfo{title}{Breakage, coalescence and size distribution of
  surfactant-laden droplets in turbulent flow},
\newblock \bibinfo{journal}{Journal of Fluid Mechanics} \bibinfo{volume}{881}
  (\bibinfo{year}{2019}) \bibinfo{pages}{244--282}.
\bibitem[{Soligo et~al.(2020)Soligo, Roccon, and Soldati}]{soligo2020effect}
\bibinfo{author}{G.~Soligo}, \bibinfo{author}{A.~Roccon},
  \bibinfo{author}{A.~Soldati},
\newblock \bibinfo{title}{Effect of surfactant-laden droplets on turbulent flow
  topology},
\newblock \bibinfo{journal}{Physical Review Fluids} \bibinfo{volume}{5}
  (\bibinfo{year}{2020}) \bibinfo{pages}{073606}.
\bibitem[{Jain and Mani(2023)}]{jain2023scalar}
\bibinfo{author}{S.~S. Jain}, \bibinfo{author}{A.~Mani},
\newblock \bibinfo{title}{A computational model for transport of immiscible
  scalars in two-phase flows},
\newblock \bibinfo{journal}{Journal of Computational Physics}
  \bibinfo{volume}{475} (\bibinfo{year}{2023}) \bibinfo{pages}{111843}.
\bibitem[{Mirjalili et~al.(2022)Mirjalili, Jain, and
  Mani}]{mirjalili2022computational}
\bibinfo{author}{S.~Mirjalili}, \bibinfo{author}{S.~S. Jain},
  \bibinfo{author}{A.~Mani},
\newblock \bibinfo{title}{A computational model for interfacial heat and mass
  transfer in two-phase flows using a phase field method},
\newblock \bibinfo{journal}{International Journal of Heat and Mass Transfer}
  \bibinfo{volume}{197} (\bibinfo{year}{2022}) \bibinfo{pages}{123326}.
\bibitem[{Chiu and Lin(2011)}]{chiu2011conservative}
\bibinfo{author}{P.-H. Chiu}, \bibinfo{author}{Y.-T. Lin},
\newblock \bibinfo{title}{A conservative phase field method for solving
  incompressible two-phase flows},
\newblock \bibinfo{journal}{Journal of Computational Physics}
  \bibinfo{volume}{230} (\bibinfo{year}{2011}) \bibinfo{pages}{185--204}.
\bibitem[{Olsson and Kreiss(2005)}]{olsson2005conservative}
\bibinfo{author}{E.~Olsson}, \bibinfo{author}{G.~Kreiss},
\newblock \bibinfo{title}{A conservative level set method for two phase flow},
\newblock \bibinfo{journal}{Journal of Computational Physics}
  \bibinfo{volume}{210} (\bibinfo{year}{2005}) \bibinfo{pages}{225--246}.
\bibitem[{Desjardins et~al.(2008)Desjardins, Moureau, and
  Pitsch}]{Desjardins2008}
\bibinfo{author}{O.~Desjardins}, \bibinfo{author}{V.~Moureau},
  \bibinfo{author}{H.~Pitsch},
\newblock \bibinfo{title}{An accurate conservative level set/ghost fluid method
  for simulating turbulent atomization},
\newblock \bibinfo{journal}{J. Comput. Phys.} \bibinfo{volume}{227}
  (\bibinfo{year}{2008}) \bibinfo{pages}{8395--8416}.
\bibitem[{Jain et~al.(2020)Jain, Mani, and Moin}]{jain2020conservative}
\bibinfo{author}{S.~S. Jain}, \bibinfo{author}{A.~Mani},
  \bibinfo{author}{P.~Moin},
\newblock \bibinfo{title}{A conservative diffuse-interface method for
  compressible two-phase flows},
\newblock \bibinfo{journal}{J. Comput. Phys.} \bibinfo{volume}{418}
  (\bibinfo{year}{2020}) \bibinfo{pages}{109606}.
\bibitem[{Jain et~al.(2023)Jain, Adler, West, Mani, Moin, and
  Lele}]{jain2023assessment}
\bibinfo{author}{S.~S. Jain}, \bibinfo{author}{M.~C. Adler},
  \bibinfo{author}{J.~R. West}, \bibinfo{author}{A.~Mani},
  \bibinfo{author}{P.~Moin}, \bibinfo{author}{S.~K. Lele},
\newblock \bibinfo{title}{Assessment of diffuse-interface methods for
  compressible multiphase fluid flows and elastic-plastic deformation in
  solids},
\newblock \bibinfo{journal}{Journal of Computational Physics}
  \bibinfo{volume}{475} (\bibinfo{year}{2023}) \bibinfo{pages}{111866}.
\bibitem[{Hwang and Jain(2023)}]{hwang2023robust}
\bibinfo{author}{H.~Hwang}, \bibinfo{author}{S.~S. Jain},
\newblock \bibinfo{title}{A robust phase-field method for two-phase flows on
  unstructured grids},
\newblock \bibinfo{journal}{arXiv preprint arXiv:2310.10795}
  (\bibinfo{year}{2023}).
\bibitem[{Brown et~al.(2023)Brown, Jain, and Moin}]{brown2023phase}
\bibinfo{author}{L.~Brown}, \bibinfo{author}{S.~Jain},
  \bibinfo{author}{P.~Moin}, \bibinfo{title}{A Phase Field Model for Simulating
  the Freezing of Supercooled Liquid Droplets}, \bibinfo{type}{Technical
  Report}, SAE Technical Paper, \bibinfo{year}{2023}.
\bibitem[{Scapin et~al.(2022)Scapin, Shahmardi, Chan, Jain, Mirjalili, Pelanti,
  and Brandt}]{scapin2022}
\bibinfo{author}{N.~Scapin}, \bibinfo{author}{A.~Shahmardi},
  \bibinfo{author}{H.~R. Chan, W}, \bibinfo{author}{S.~Jain, S},
  \bibinfo{author}{S.~Mirjalili}, \bibinfo{author}{M.~Pelanti},
  \bibinfo{author}{L.~Brandt},
\newblock \bibinfo{title}{A mass-conserving pressure-based method for two-phase
  flows with phase change},
\newblock \bibinfo{journal}{Proceedings of the Summer Program}
  (\bibinfo{year}{2022}) \bibinfo{pages}{195--204}.
\bibitem[{Collis et~al.(2022)Collis, Mirjalili, Jain, and Mani}]{collis2022}
\bibinfo{author}{H.~Collis}, \bibinfo{author}{S.~Mirjalili},
  \bibinfo{author}{S.~Jain, S}, \bibinfo{author}{A.~Mani},
\newblock \bibinfo{title}{Assessment of weno and teno schemes for the four
  equation- compressible two-phase flow model with regularization term},
\newblock \bibinfo{journal}{Annual Research Briefs}  (\bibinfo{year}{2022})
  \bibinfo{pages}{151--165}.
\bibitem[{Liang et~al.(2023)Liang, Wang, Wei, and Xu}]{liang2023lattice}
\bibinfo{author}{H.~Liang}, \bibinfo{author}{R.~Wang},
  \bibinfo{author}{Y.~Wei}, \bibinfo{author}{J.~Xu},
\newblock \bibinfo{title}{Lattice boltzmann method for interface capturing},
\newblock \bibinfo{journal}{Physical Review E} \bibinfo{volume}{107}
  (\bibinfo{year}{2023}) \bibinfo{pages}{025302}.
\bibitem[{Jain and Moin(2022)}]{jain2022kinetic}
\bibinfo{author}{S.~S. Jain}, \bibinfo{author}{P.~Moin},
\newblock \bibinfo{title}{A kinetic energy--and entropy-preserving scheme for
  compressible two-phase flows},
\newblock \bibinfo{journal}{Journal of Computational Physics}
  \bibinfo{volume}{464} (\bibinfo{year}{2022}) \bibinfo{pages}{111307}.

\end{thebibliography}

\end{document}